\documentclass[12pt]{iopart}
\usepackage{amssymb}
\usepackage{iopams}
\usepackage{epsf}

\begin{document}

\noindent 
{\scriptsize Published in Journal of Physics G: Nuclear and Particle
Physics. Copyright 1999 IOP Publishing Ltd}

\title{Electromagnetic showers in a strong magnetic field}

\author{V Anguelov and H Vankov}

\address{Institute for Nuclear Research and Nuclear Energy, Bulgarian 
Academy of Sciences, 72 Tzarigradsko Chaussee, 1784 Sofia, Bulgaria}

\begin{abstract}
We present the results concerning the main shower characteristics in a
strong magnetic field obtained through shower simulation. The processes of
magnetic bremsstrahlung and pair production were taken into account for
values of the parameter $\chi \gg 1$. We compare our simulation results with
a recently developed cascade theory in a strong magnetic field.
\end{abstract}

\section{Introduction}

Electromagnetic showers are a universal phenomenon. Besides occurring in
matter or radiation field cascade, multiplication of electrons and photons
can arise in a strong magnetic field. Such super-strong fields ($\sim
10^{12}\ \mathrm{G}$) probably exist in the vicinity of some astrophysical
objects such as pulsars, for example. In this case rotating neutron stars
induce strong electric fields above the polar cap. Accelerated by these
fields, high-energy particles (with energy up to $\sim 10\ \mathrm{TeV}$)
move along curved magnetic field lines and emit curvature photons. The
energy of these photons is enough to produce electron-positron pairs in
magnetic and electric fields. The subsequent quantized synchrotron radiation
by pairs will convert to a second generation of pairs and then an
electromagnetic cascade develops in the pulsar magnetosphere. The shower
development determines, to a considerable extent, the properties of the
observed radiation from these objects. This was proposed for the first time
in \cite{sturrock}. Since electromagnetic cascades in strong magnetic fields
were considered in many works mainly in connection with specific models of
radio pulsars, gamma-ray bursts, blazars (see, for example, \cite{harding}-%
\cite{ bednarek}).

It is well known that the essentially non-zero probabilities for magnetic
bremsstrahlung and pair production require both strong field and high
energies \cite{erber}. The relevant parameter determining the criteria for
this is:

\[
\chi =\frac \varepsilon {mc^2}\frac H{H_{cr}} 
\]
where $\varepsilon $ is the particle energy, $H$ is the magnetic field
strength, $m$ is the electron mass and $H_{cr}=4.41\times 10^{13}\ \mathrm{G}
$.

\begin{figure}[tbp]
\begin{center}
\mbox{\epsfysize=2.4in \epsffile{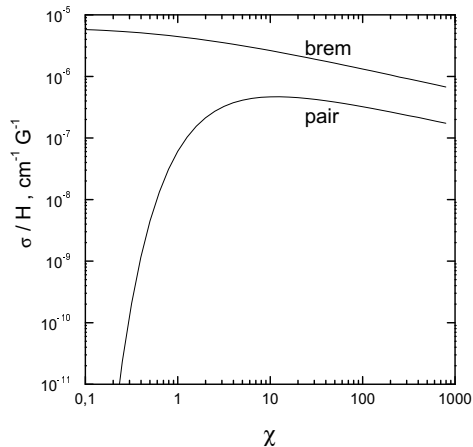}}
\end{center}
\caption{The total cross sections for magnetic bremsstrahlung and pair
production as function of $\chi $.}
\end{figure}

The total probabilities (cross sections) for radiation and pair production
for a given value of the magnetic field strength depend only on $\chi $ and
are shown in figure 1. One can see that magnetic pair production has
significant probability for $\chi \gtrsim 0.1$ (photon energy must be $%
\gtrsim 2mc^2$). For effective shower development one needs even higher
values of $\chi $ ($\chi \gtrsim 1$) because with increasing $\chi $ the
radiated photon spectrum becomes harder. For $\chi \gtrsim 1$ (quantum
region) the energy of the radiated photon is of the order of the electron
energy. It is interesting to note that for a photon with energy $\gtrsim
7,5\times 10^{19}\ \mathrm{eV}$ even Earth's magnetic field ($\sim 0.3\ 
\mathrm{G}$) is strong enough ($\chi \sim 1$) to be a good environment for
creating an electromagnetic shower. If such extremely high-energy photons
are presented in the primary cosmic ray flux they will undergo cascading in
the geomagnetic field before entering the Earth's atmosphere. This problem
has been intensively discussed recently in connection with the detection of
the highest cosmic ray events \cite{halzen}.

As mentioned above, most of the treatments of cascades in a magnetic field
were connected with specific models of astrophysical objects and are
numerical in nature. The most general treatment of cascade properties
emphasizing an analytical approach is made in \cite{baring}, where the
steady-state kinetic equations for the electron-positron and photon
distributions are solved in a strong magnetic field.

A different approach to the shower study in a magnetic field is applied in 
\cite{kanevsky}. It is similar to those for showers in matter and was
motivated by the study of the primary gamma rays with extremely high
energies ($\gtrsim 10^{20}\ \mathrm{eV}$) propagating through the
geomagnetic field and Earth's atmosphere. The average shower characteristics
obtained by numerically solving the system of cascade equations, show some
of the main features of the cascade. While the shower is similar to those in
matter for $\chi \gg 1$, its nature changes sharply for $\chi \lesssim 1$
which is connected with sharp increase of the photon free path.

Recently, a kinetic theory of electromagnetic showers in a strong magnetic
field has been developed in a similar to the cascade theory in matter, in
approximation A \cite{akhiezer}. Electromagnetic shower theories have been
developed since 1937 following the works of Bhabha and Heitler and Carlson
and Oppenheimer \cite{bhabha}. Landau and Rumer \cite{landau} developed a
complete theory in approximation A. Their work contains the formalism which
is widely used in later shower theories.

In the shower theory the mathematical description of the cascade process is
based on the Boltzmann kinetic equation for particle flux density. The
system of integro-differential equations of the one-dimensional Landau-Rumer
theory is universal because it describes the shower development in any
substance. Only the expressions for pair creation and bremsstrahlung
probabilities per unit length are different. Using asymptotic forms of both
processes for very high energies in a strong magnetic field, analytic
formulae similar to those of standard cascade theory in approximation A for
one-dimensional shower characteristics are obtained in \cite{akhiezer}. But,
as in matter, the kinetic equations were solved within certain
approximations.

In this work we give the results from the Monte Carlo simulation of the
longitudinal development of electromagnetic showers in a strong magnetic
field for $\chi \gg 1$. We present shower profiles for different ratios of
the primary and threshold energies, $E_0/E$, and energy spectra of shower
particles at different depths. We analyse the behaviour of the shower
maximum with respect to $E_0/E$. We compare our modelled results with
theoretical ones in order to estimate theoretical approximations and the
range of their validity.

\section{The probability functions}

The main elementary processes leading to particle multiplication in a
magnetic field are magnetic bremsstrahlung and magnetic pair production. The
corresponding probabilities per unit length are \cite{akhiezer,bayer}:

\begin{eqnarray}
\fl \pi \left( \varepsilon ,\omega \right) \rmd\omega =\frac{\alpha m^2}{%
\pi \sqrt{3}}\frac{\rmd\omega }{\varepsilon ^2}\left[ \left( \frac{%
\varepsilon -\omega }\varepsilon +\frac \varepsilon {\varepsilon -\omega
}\right) K_{\frac 23}\left( \frac{2u}{3\chi }\right) -\int\nolimits_{\frac{2u%
}{3\chi }}^\infty K_{\frac 13}\left( y\right) \rmd y\right]  \nonumber \\
&&  \label{eq1} \\
\fl \gamma \left( \omega ,\varepsilon \right) \rmd\varepsilon =\frac{%
\alpha m^2}{\pi \sqrt{3}}\frac{\rmd\varepsilon }{\omega ^2}\left[ \left( 
\frac{\omega -\varepsilon }\varepsilon +\frac \varepsilon {\omega
-\varepsilon }\right) K_{\frac 23}\left( \frac{2u_1}{3\chi }\right)
+\int\nolimits_{\frac{2u_1}{3\chi }}^\infty K_{\frac 13}\left( y\right) \rmd %
y\right]  \nonumber
\end{eqnarray}
where $\varepsilon $ and $\omega $ are the electron and photon energy and $%
u=\frac \omega {\varepsilon -\omega }$,\ $u_1=\frac{\omega ^2}{\varepsilon
(\omega -\varepsilon )}$. Parameter $\chi $ was defined above. Here $\hbar
=c=1$. $K_\nu \left( z\right) =\int\nolimits_0^\infty \rme^{-z\ \mathrm{ch}%
(t)}\mathrm{ch}\left( \nu t\right) \rmd t$ is a modified Bessel function
known as MacDonald's function. Parameter $\nu $ can have any\ real or
complex values, here $\nu =\frac 23$ and $\frac 13.$ For simplicity we
assume that the electron (positron) is moving perpendicular to the magnetic
field $H$. As already mentioned, the probabilities of both processes are
essentially different from zero under the condition that the parameter $\chi
\gg 1$ , which means that the particle energy $\varepsilon >>\varepsilon _c$%
, where $\varepsilon _c=mc^2\frac{H_{cr}}H$. It is mentioned in \cite{bayer}
that this condition corresponds to the conditions $u/\chi <<1$ and $u_1/\chi
<<1$, i.e. one can use in (\ref{eq1}) the asymptotic form of $K_\nu \left(
z\right) $ for $z<<1$, $K_\nu (z\ll 1)=\frac{\Gamma (\nu )}{2^{1-\nu }}%
(\frac 1z)^\nu +...$. Then expressions (\ref{eq1}) may be simplified:

\begin{eqnarray}
\pi \left( \varepsilon ,\omega \right) &=&q\left[ \frac{\left( \varepsilon
-\omega \right) ^{\frac{5}{3}}}{\varepsilon ^{\frac{7}{3}}\omega ^{\frac{2}{3%
}}}+\frac{1}{\left( \varepsilon -\omega \right) ^{\frac{1}{3}}\varepsilon ^{%
\frac{1}{3}}\omega ^{\frac{2}{3}}}\right] ,  \nonumber \\
&&  \label{eq2} \\
\gamma \left( \omega ,\varepsilon \right) &=&q\left[ \frac{\left( \omega
-\varepsilon \right) ^{\frac{5}{3}}}{\omega ^{\frac{8}{3}}\varepsilon ^{%
\frac{1}{3}}}+\frac{\varepsilon ^{\frac{5}{3}}}{\omega ^{\frac{8}{3}}\left(
\omega -\varepsilon \right) ^{\frac{1}{3}}}\right] ,  \nonumber
\end{eqnarray}
where $q=3.9\times 10^{6}\left[ \frac{H}{H_{c}}\right] ^{\frac{2}{3}}\frac{%
\mathrm{GeV}^{\frac{1}{3}}}{\mathrm{cm}}$. Using $u=\omega /\varepsilon $
(correspondingly $u=\varepsilon /\omega $) we can rewrite (\ref{eq2}) in the
form:

\begin{eqnarray}
\pi \left( \varepsilon ,\omega \right) d\omega &=&\frac{q}{\varepsilon ^{%
\frac{1}{3}}}\left[ \frac{\left( 1-u\right) ^{\frac{5}{3}}}{u^{\frac{2}{3}}}+%
\frac{1}{\left( 1-u\right) ^{\frac{1}{3}}u^{\frac{2}{3}}}\right] \rmd u, 
\nonumber \\
&&  \label{eq3} \\
\gamma \left( \omega ,\varepsilon \right) d\varepsilon &=&\frac{q}{\omega ^{%
\frac{1}{3}}}\left[ \frac{\left( 1-u\right) ^{\frac{5}{3}}}{u^{\frac{1}{3}}}+%
\frac{u^{\frac{5}{3}}}{\left( 1-u\right) ^{\frac{1}{3}}}\right] \rmd u. 
\nonumber
\end{eqnarray}

The total probabilities per unit length for bremsstrahlung and pair
production are given by

\begin{eqnarray}
W_r\left( \varepsilon \right) &=&\int\nolimits_0^\varepsilon \pi \left(
\varepsilon ,\omega \right) \rmd\omega =\frac q{\varepsilon ^{\frac
13}}\int\nolimits_0^1\left[ \frac{\left( 1-u\right) ^{\frac 53}}{u^{\frac 23}%
}+\frac 1{\left( 1-u\right) ^{\frac 13}u^{\frac 23}}\right] \rmd u
\label{eq4} \\
\ &=&5.642\frac q{\varepsilon ^{\frac 13}},  \nonumber \\
W_p\left( \omega \right) &=&\int\nolimits_0^\omega \gamma \left( \omega
,\varepsilon \right) \rmd\varepsilon =1.467\frac q{\omega ^{\frac 13}}.
\label{eq5}
\end{eqnarray}

Unlike the Bethe-Heitler probability,$\ \pi (\varepsilon ,\omega )$ does not
contain an infrared divergence and because of this $W_{r}\left( \varepsilon
\right) $ is finite.

As mentioned earlier, the probabilities of both processes were obtained
using the asymptotic form of the function $K_\nu \left( z\right) $ for $z<<1$%
. This explains the behaviour of the total cross sections as a function of
the particle energy of power $(-\frac 13)$ (at fixed $H$) and in figure 1
this is the region of $\chi \gg 1$. Expression (\ref{eq5}) coincides with
the expression of the photon attenuation coefficient given in Erber's review 
\cite{erber} when the asymptotic form of auxiliary function $T(\chi )$ for $%
\chi >>1$, $T(\chi )\sim 0.60\chi ^{\frac 13}$, is used. Here $\chi =\frac 12%
\frac{h\nu }{mc^2}\frac H{H_{cr}}$, $h\nu $ is the photon energy.

\section{Simulation}

To investigate the shower characteristics in a strong magnetic field we
developed our own Monte Carlo code. The probabilities (\ref{eq3}) were used
to sample energies of the secondary particles - photon in bremsstrahlung and
electron in pair creation. We constructed tables with cumulative
distributions from probabilities (\ref{eq3}) through a small step by $u$.
The mean interaction length is the inverse value of the total probability, (%
\ref{eq4}) and (\ref{eq5}).

It was shown in \cite{akhiezer} that the average ranges of both electron and
photon, with energy $E_0$ in a magnetic field $H$, are of the same order of
magnitude, $\sim \frac{E_0^{\frac 13}}q$. The quantity $L=\frac{E_0^{\frac
13}}q$ which includes the matter parameters (magnetic field strength $H$)
could play the role of a radiation length. In this case $E_0$ is the energy
of the particle initiating the shower. It is important to note, that unlike
matter, here $L$ depends on the energy $E_0$ of the primary particle. $L$ is
a relatively small quantity. For example, in magnetic field $H=10^5\ \mathrm{%
G}$ \ for $E_0=10^6\ \mathrm{GeV}$,\ $L=14.86\ \mathrm{cm}$, for $E_0=10^9\ 
\mathrm{GeV}$,\ $L=148.6\ \mathrm{cm}$.

We considered the problem as one dimensional, i.e. we assumed that the
shower is only developing in the direction of the primary particle entering
the magnetic field at $t=0$. The distance is measured in units $L$. All
shower particles were followed down to some threshold energy $E$.

\begin{figure}[tbp]
\begin{center}
\mbox{\epsfysize=2.4in \epsffile{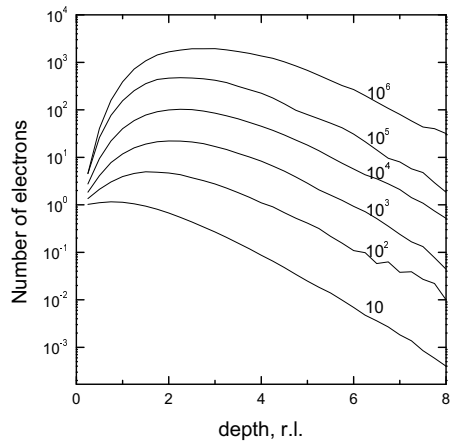}}
\end{center}
\caption{Shower profiles for electron induced showers. Numbers refer to the
ratio $E_{0}/E$.}
\end{figure}

\begin{figure}[tbp]
\begin{center}
\mbox{\epsfysize=2.4in \epsffile{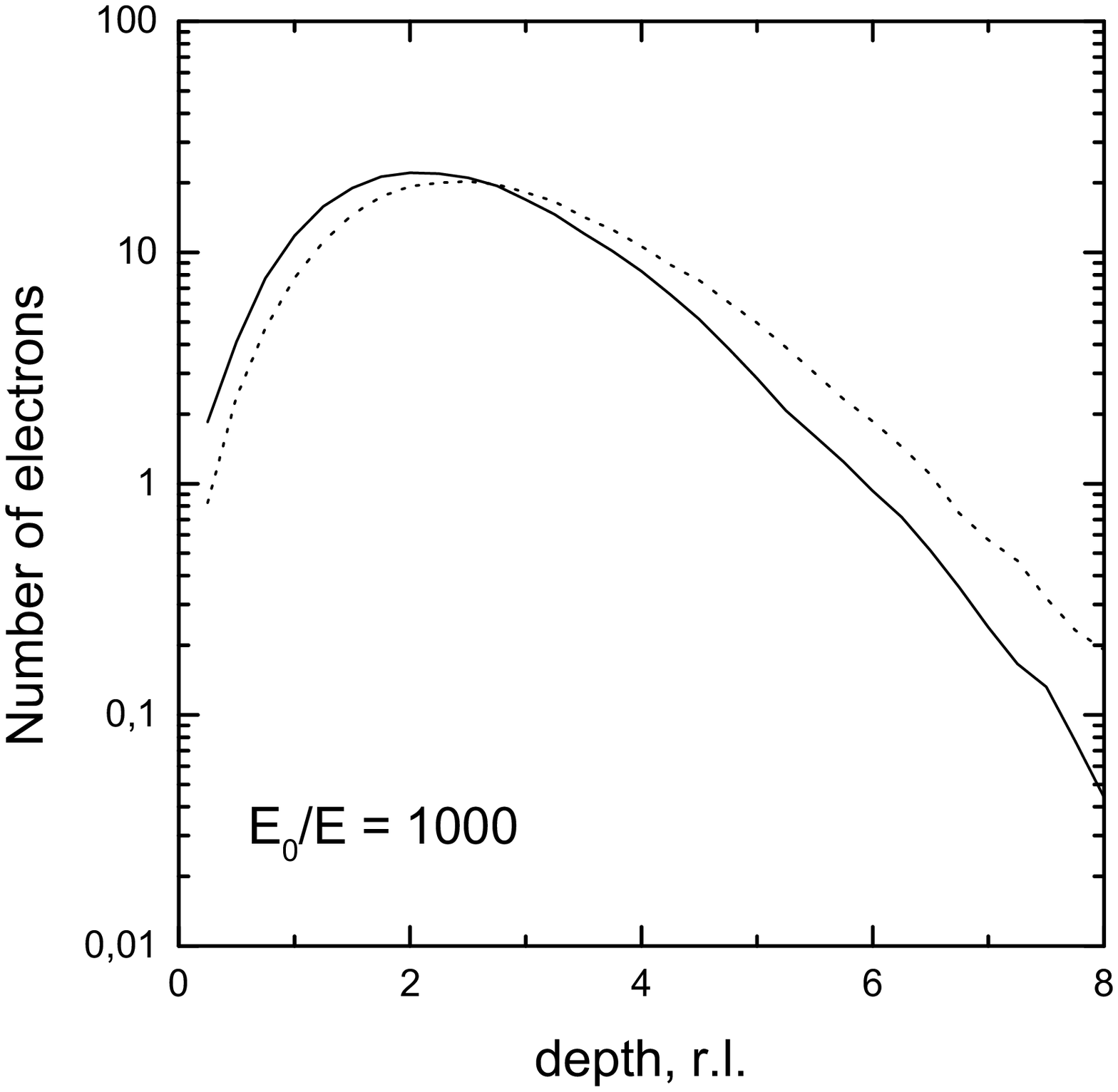}}
\end{center}
\caption{Shower profiles for electron (solid curve) and photon (dashed
curve) induced showers.}
\end{figure}

\section{Results and discussion}

We performed simulation for various sets of primary and threshold energies ($%
E_{0}$ and $E$) and different magnetic field strengths $H.$

Our results confirmed the theoretical prediction that the above-mentioned
quantity, $L$, plays the role of a radiation length. Similar to the standard
shower theory under approximation A, the longitudinal cascade development is
independent of an absorber (magnetic field) when distances are measured in
radiation lengths and the average behaviour of a shower is expressed by a
function of $E_0/E.$

Shower profiles for electron-initiated showers and different $E_0/E$ are
shown in figure 2. The corresponding curves for photons are very close to
those of electrons and because of this they are not shown. When the primary
particle is a photon (figure 3), the shower maximum is shifted by $\sim 
\frac{1}{2}\ \mathrm{r.l.}$ (where r.l. denotes radiation length) deeper,
which is close to the difference between the mean interaction lengths of
primary electron and photon.

The differential energy spectra of shower particles at different stages of
the shower development are shown in figure 4. The depth of $2.25\ \mathrm{r.l%
}$. is near the shower maximum. The spectra can be described by power-law of
energy, $E\frac{\rmd n}{\rmd E}\sim E^{-\delta }$, with $\delta $ increasing
with the depth and approaching $1$ for $t\rightarrow \infty $.

As can be seen, the typical distance over which the shower develops is a few
radiation lengths. The depth of the shower maximum increases very slowly
with the increase of $E_{0}/E$ approaching a limit. In very strong fields,
e.g. in pulsars, $L$ becomes very small ($\sim 10^{-4}\ \mathrm{cm}$ for $%
E_{0}=10^{6}\ \mathrm{GeV}$) which means that the shower has the
longitudinal spread of the same order. This confirms the theoretical
prediction in \cite{akhiezer} that the strong magnetic fields are effective
screens for very high energy electrons, positrons and photons.

\begin{figure}[tbp]
\begin{center}
\mbox{\epsfysize=2.4in \epsffile{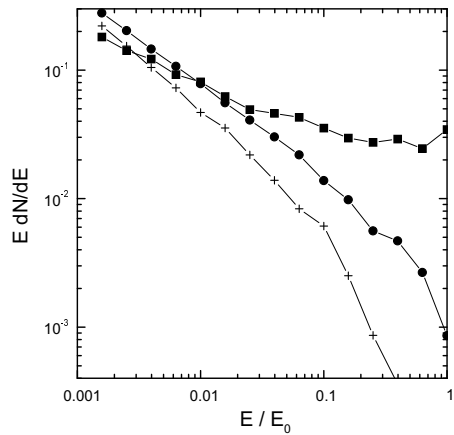}}
\end{center}
\caption{Energy spectrum of electrons at three depths: line with squares-$%
0.75\ \mathrm{r.l.}$, line with circles-$2.25\ \mathrm{r.l}$. (near the
shower maximum), line with crosses-$5\ \mathrm{r.l.}$ $E_{0}/E=1000$.}
\end{figure}

\begin{figure}[tbp]
\begin{center}
\mbox{\epsfysize=2.4in \epsffile{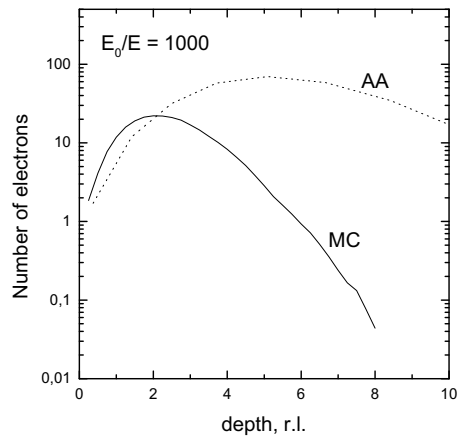}}
\end{center}
\caption{Simulated shower profile (MC) compared with the theoretical curve
in the adiabatic approximation (AA).}
\end{figure}

\begin{figure}[tbp]
\begin{center}
\mbox{\epsfysize=2.4in \epsffile{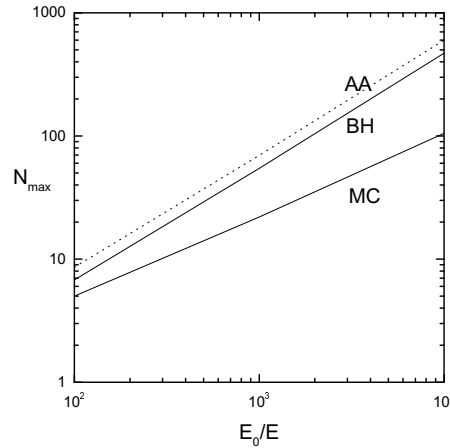}}
\end{center}
\caption{The number of electrons in the shower maximum as a function of $%
E_{0}/E$. MC-this work, AA-adiabatic approximation and BH-classical cascade
theory in approximation A.}
\end{figure}

\begin{figure}[tbp]
\begin{center}
\mbox{\epsfysize=2.4in \epsffile{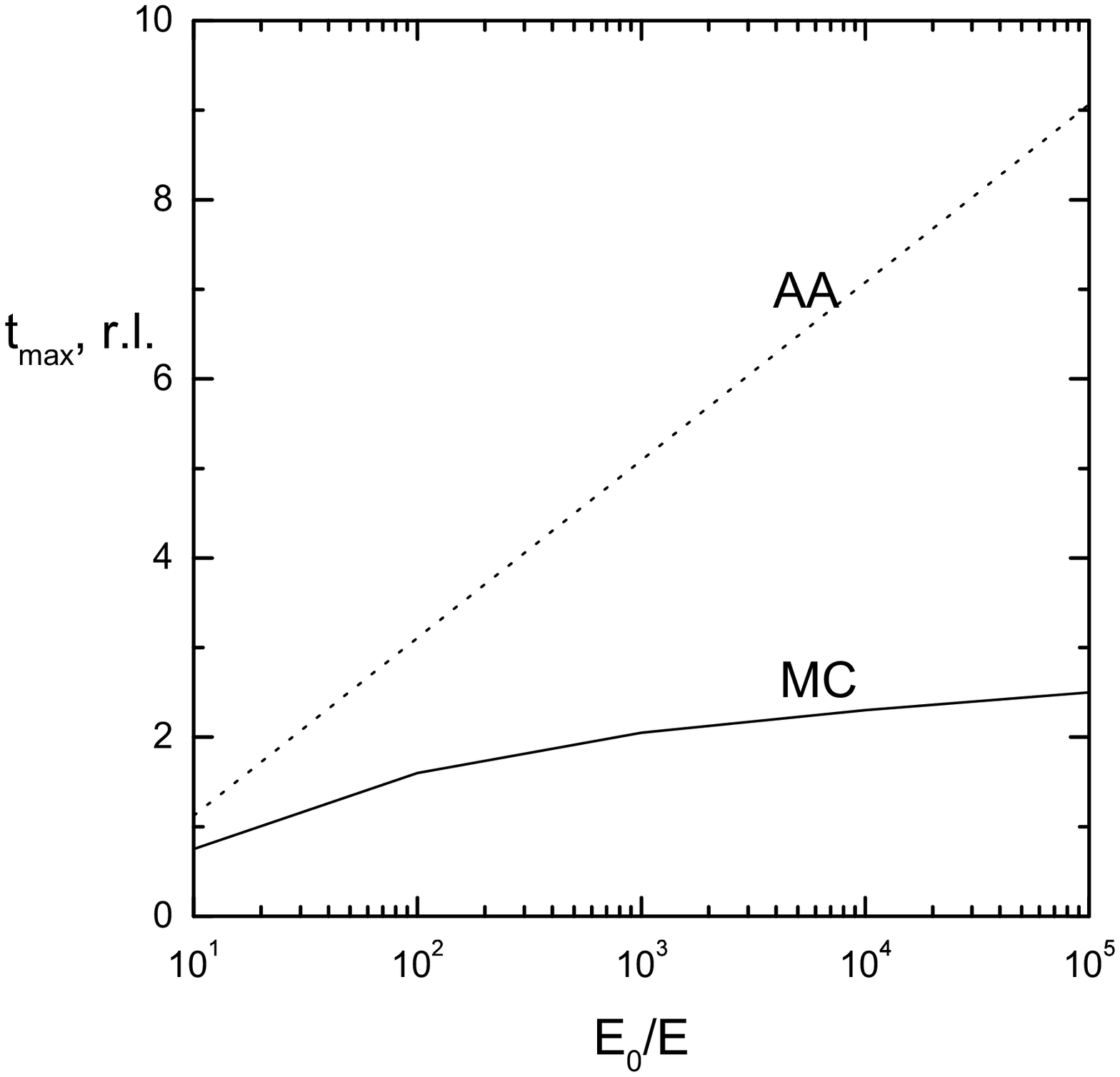}}
\end{center}
\caption{Depth of the maximum as a function of $E_{0}/E$.}
\end{figure}

It should be noted here that these properties of the shower are valid for
such values of $E_{0}$, $E$ and $H$ that satisfy the condition $\chi \gg 1$
and the asymptotic expressions (2) can also be used. In addition, to
consider the shower as one-dimensional, the electron energies must obey the
condition \cite{akhiezer}

\[
\left( \frac{\varepsilon }{mc^{2}}\right) ^{\frac{2}{3}}\left( \frac{H_{cr}}{%
H}\right) ^{\frac{1}{3}}\frac{\alpha }{2\pi }\gg 1. 
\]

This criteria comes from the requirement that the electron gyroradius $%
R=\varepsilon /eH$ must be much greater than the typical length of the
shower, i.e. $R\gg L$.

However, detailed comparison of a numerical results shows that the Monte
Carlo cascade curves differ significantly from the theoretical ones. This is
illustrated in figure 5 where shower profiles for the electron-induced
showers and $E_0/E=1000$ are shown.

This disagreement could probably be explained with the approximations used
in \cite{akhiezer} to get a solution of cascade equations. In the case of a
magnetic field, the shower theory is more complicated than the conventional
theory in approximation A because the probabilities $\pi \left( \varepsilon
,\omega \right) \rmd\omega $ and $\gamma \left( \omega ,\varepsilon \right) %
\rmd\varepsilon $ are not scaling functions, i.e. they do not depend only on
the ratio of secondary to primary energies (functions of $\omega
/\varepsilon \;$or $\varepsilon /\omega $ in our notation). Mellin
transforms lead to differential-difference equations for the distribution
functions and its solutions are found by authors of \cite{akhiezer} in the
so-called adiabatic and modified adiabatic approximations. As pointed out in 
\cite{akhiezer}, the adiabatic approximation (AA) is limited to the region
after the maximum. In the modified adiabatic approximation it is assumed
that the explicit dependence of $\pi \left( \varepsilon ,\omega \right) $
and $\gamma \left( \omega ,\varepsilon \right) $ on primary energy can be
eliminated by its substitution with some mean energy per interaction in the
shower. This leads to the same solutions with the modified radiation length
but, obviously, the shower development is distorted. If we use the
probabilities $\pi \left( \varepsilon ,\omega \right) $ and $\gamma \left(
\omega ,\varepsilon \right) $ in this modified way in our Monte Carlo code
we obtain results very close to the theoretical ones. Figure 6 shows the
number of electrons in the shower maximum $N_{\max }$ as function of $E_0/E$%
. One can see that the rise of $N_{\max }$ with $E_0/E$ is slower than that
of AA of the theory. $N_{\max }$ is exactly proportional to $\left(
E_0/E\right) ^\xi $ . AA gives $\xi $ close to one as it is in the standard
theory with Bethe-Heitler cross sections (curve labelled BH in the figure)
while our simulation (curve labelled MC) gives $\xi \approx \frac 23$.
Figure 7 shows the depth of the maximum $t_{\max }$ as a function of $E_0/E$
for both theoretical and simulated showers. As can be seen, the behaviour of
the modelled $t_{\max }$ is too different from those of logarithmically
increasing theoretical $t_{\max }$. After some $E_0/E$ the modelled $t_{\max
}$ almost ceases to increase.

It is easy to demonstrate this feature of showers in a magnetic field using
Heitler's elementary cascade model \cite{rossi}. In this model,
electromagnetic particles subdivide into two particles with half the initial
energy. In matter this takes place on every radiation length. In a magnetic
field, however, the situation is substantially different. After each
interaction the interaction length, which is a function of energy power of $%
\frac{1}{3}$, decreases due to particle energy splitting. If the primary
particle has an energy $E_0$ and the interaction length $t_0\ \mathrm{r.l.}$%
, then after the first interaction the number of particles will be two, each
of energy $E_0/2$. The next interaction length will be $t_0/2^{\frac 13}$
and the number of particles four, each of energy $E_0/4$. After $N$
interactions the particle number is already $2^N$ and their energy $%
E=E_0.2^{-N}$. The next interaction length will be $\frac{t_0}{\left[
2^{\frac 13}\right] ^N}\ \mathrm{r.l.}$ The distance at which this takes
place is

\begin{figure}[tbp]
\begin{center}
\mbox{\epsfysize=2.4in \epsffile{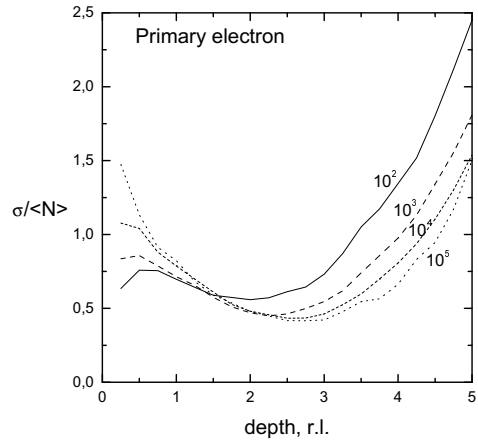}}
\end{center}
\caption{Fluctuations of the number of electrons $\left( \sigma /\langle
N\rangle \right) $ as a function of the depth for different $E/E_{0}$.}
\end{figure}

\begin{figure}[tbp]
\begin{center}
\mbox{\epsfysize=2.4in \epsffile{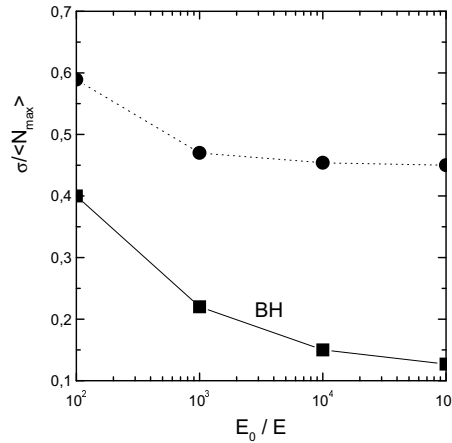}}
\end{center}
\caption{Fluctuations of the number of electrons in the shower maximum $%
\left( \sigma /\langle N\rangle \right) $ as a function of $E/E_{0}$. Dotted
line-showers in a magnetic field, solid line-BH showers.}
\end{figure}

\begin{eqnarray*}
\fl T =t_0+\frac{t_0}{2^{\frac 13}}+\frac{t_0}{\left( 2^{\frac 13}\right) ^2}%
+...+\frac{t_0}{\left( 2^{\frac 13}\right) ^{N-1}}=t_0\left[ 1+\frac
1{2^{\frac 13}}+\frac 1{\left( 2^{\frac 13}\right) ^2}+...+\frac 1{\left(
2^{\frac 13}\right) ^{N-1}}\right] . \\
&&
\end{eqnarray*}

The expression in the square brackets is a sum of a geometrical progression
with $q=2^{-1/3}$ and thus

\begin{eqnarray}
T &=&4.847t_0\left[ 1-\frac 1{\left( 2^{\frac 13}\right) ^{N-1}}\right] .
\label{eq6} \\
&&  \nonumber
\end{eqnarray}
\ This expression shows that if $N$ increases, i.e. $E_0/E$ increases, $%
t_{\max }$ approaches a limit. This means that after some $N$ (or $E_0/E$) $%
\ t_{\max }$ practically ceases to increase. In our rather simplified model
the maximum $t_{\max }\ $is $4.847t_0\ \mathrm{r.l.}$ The first interaction
length $t_0$ is $0.117\ \mathrm{r.l}$. for the electron and $0.682\ \mathrm{%
r.l.}$ for the photon. If we take the greater value this will lead to an
estimation of the maximum $t_{\max }$ of $3.31\ \mathrm{r.l.}$ whose value
does not contradict with the curve modelled in figure 7.

Another important characteristic of electromagnetic showers in a strong
magnetic field which can be easily obtained from the simulation are
fluctuations in the shower development. There is no theoretical
consideration of this problem. The problem is complicated even for the
standard cascade theory (see, e.g. \cite{uchaikin}).

In figure 8 the fluctuations of the number of shower electrons $\left(
\sigma /\langle N\rangle \right) $ as a function of the depth for different $%
E_0/E$ are shown. The primary particle is an electron. The behaviour of the
curves is very similar to those of BH showers but the fluctuations in a
magnetic field are significantly larger. This is illustrated in figure 9
where fluctuations in the shower maximum as a function of $E_0/E$ are shown,
compared with BH showers. Calculations for BH showers are performed by
direct MC simulation in air for very high energies $E_0$ and $E$, i.e. at
the conditions where the standard cascade theory in approximation A is valid.

The larger fluctuations in a magnetic field compared to BH showers are not
an unexpected result. The main sources for the fluctuations of the number of
particles in the cascade are fluctuations of interaction lengths and the
random energy distribution of created secondary particles. Unlike the BH
cross sections the mean interaction lengths for magnetic bremsstrahlung and
pair-production processes are functions of the particle energy of power $%
\frac 13$ (see (\ref{eq4},\ref{eq5})). The differential cross sections (\ref
{eq3}) depend on the particle energy as well. Their features lead to a
significant probability that the secondary electron or photon will take a
large fraction of the primary energy.

\section{Conclusions}

A direct MC simulation of the longitudinal development of electromagnetic
showers in a strong magnetic field has been performed. The processes of the
magnetic bremsstrahlung and of the magnetic pair production were included in
the simulation with asymptotic expressions for both probabilities valid for
very high energies. Simulated results were compared with the theory of
showers in a strong magnetic field developed in \cite{akhiezer}. The main
predictions of this theory - the dependence of the radiation length on
magnetic field strength and the energy of the primary particle, the very
small shower longitudinal spread and the closeness of electron and photon
profiles, are confirmed by our simulation.

However, the AA of the theory used in \cite{akhiezer} was inadequate for a
precise qualitative estimate of shower characteristics.

\section*{Acknowledgments}

The authors thanks A. Rekalo and T.Stanev for helpful discussions. This work
was partially supported by a grant F-460 of the Bulgarian NFSR and by the
Bulgarian Science and Culture Foundation.


\section*{References}

\begin{thebibliography}{99}
\bibitem{sturrock}  Sturrock P A 1971 \textit{Ap.J. }\textbf{164} 529

\bibitem{harding}  Daugherty J K and Harding A 1982 \textit{Ap.J. }\textbf{%
252} 337

\bibitem{baring}  Baring M 1989 \textit{Astron. Astrophys. }\textbf{225} 260

\bibitem{bednarek}  Bednarek W 1997 \textit{Mon. Not. R. Astron. Soc.} 
\textbf{285} 69

\bibitem{erber}  Erber T 1966 \textit{Rev. Mod. Phys. }\textbf{38} 626

\bibitem{halzen}  Halzen F, Vazquez R, Stanev T and Vankov H P 1995 \textit{%
\ Astrophys. Phys.} \textbf{3} 151

\bibitem{kanevsky}  Kanevsky B L and Goncharov A I 1989 \textit{Voprosy
atomnoy nauki i techniki. Seria: Technika fiz. exp. }\textbf{4} 1 (in
Russian)

\bibitem{akhiezer}  Akhiezer A I, Merenkov N P and Rekalo A P 1994 \textit{%
J. Phys. G: Nucl. Part. Phys. } \textbf{20} 1499\\Akhiezer A I, Merenkov N P
and Rekalo A P 1995 \textit{Nucl. Phys.} \textbf{58} 491 (in Russian)

\bibitem{bhabha}  Bhabha H J and Heitler W 1937 \textit{Proc. R. Soc.} A 
\textbf{519} 432\\Carlson J F and Oppenheimer J R 1937 \textit{Phys. Rev.} 
\textbf{51} 220

\bibitem{landau}  Landau L and Rumer G 1938 \textit{Proc. R. Soc.} A \textbf{%
166} 213

\bibitem{bayer}  Bayer V H, Katkov B M and Fadin V S 1973 \textit{Radiation
of Relativistic Electrons }(Moscow: Atomizdat) (in Russian)

\bibitem{rossi}  Rossi B 1952 \textit{High Energy Particles }(New York)

\bibitem{uchaikin}  Uchaikin V V and Ryzhkov V V 1998 \textit{Stochastic
Theory of High Energy Particle Transfer} (Novosibirsk: Nauka) (in Russian)
\end{thebibliography}
\end{document}